\documentstyle[12pt]{article}
\begin{document}
\centerline{The chemical potential of the electron gas on a two
dimensional lattice}
\begin{center}
\medskip
V.Celebonovic
\end{center}
\begin{center}
Institute of Physics,Pregrevica 118,11080 Zemun-
Beograd,Yugoslavia
\end{center}
\begin{center}
e-mail: vladan@phy.bg.ac.yu
\end{center}
\newpage
Abstract: The chemical potential of the electron gas on a two-dimensional
rectangular lattice is determined.An approximate expression for $\exp(-\mu/T)$ is obtained,and its second order approximation is discussed in some detail.
\newpage
\begin{center}
Introduction
\end{center}

Investigations of correlated electron systems of various dimensionality are one
of the priorities of contemporary condensed matter physics (for example [1]).The
driving force behind the scientific interest in such systems is the fact that they exist in high temperature superconductors
and organic conductors.Although the existence of correlated electrons in these
materials is regarded as an "established fact",it is far from being completely
clear how does one theoretically describe them [2].The question is if it is possible
to use the Fermi liquid model,or the Luttinger liquid model has to be used.For a recent introduction to the Luttinger liquid theory see  for example [3]. 

The aim of the present paper is to calculate the chemical potential of the
electron gas on a two dimensional rectangular lattice.Theoretically,this is an
interesting problem in itself,and the experimental motivation is amply
discussed in the literature (for example [4]). The calculation will be
performed within the standard Fermi liquid  (FL) model.This presumption could be
criticized on the grounds that there are recent results indicating the
breakdown of the FL model in cuprate superconductors and heavy-fermion
systems(such as [5]).The idea of the
present calculation is to contribute to this result,in a different way: by calculating the chemical potential,then
separately the electrical conductivity and finally drawing conclusions by comparison
with experimental data.
\begin{center}
Calculations
\end{center}
The lattice axes are designated by x and y.It can be shown that the number density of a 2D Fermi liquid is given by [6]:
\begin{equation}
\label{(1)}\ n=\frac{<N>}{S}= \int\frac{exp(\beta(\mu-\epsilon))}{1+exp(\beta(\mu-
\epsilon))} 2 \frac{d^{2}p}{(2\pi\hbar)^{2}}
\end{equation}
where $n$ denotes the number density and all the other symbols have their usual meanings and $p=\hbar k$. It then
follows that $dp=\hbar\frac{\partial k}{\partial\epsilon}d\epsilon$,where
$\epsilon$ denotes the energy.Accordingly,
\begin{equation}
\label{(2)}\ d^{2}p = dp_{x}dp_{y}=\hbar^{2}\frac{\partial
k_{x}}{\partial\epsilon_{x}}\frac{\partial
k_{y}}{\partial\epsilon_{y}}d\epsilon_{x}d\epsilon_{y}
\end{equation}
\newpage
Assuming that the energy bands along each of the lattice axes have the form
$\epsilon=-2tcos(ks)$,where $t$ denotes the hopping integral and $s$ the
lattice constant along the axis,it follows that
\begin{equation}
\label{(3)} \frac{\partial k_{x}}{\partial\epsilon_{x}}=\frac{1}{2s_{x}t_{x}}\frac{d\epsilon_{x}}{\surd(1-(\epsilon_{x}/(2t_{x}))^{2})}
\end{equation}
An analogous relation is valid along the y axis.This implies that
\begin{equation}
\label{(4)}\ d^{2}p = \frac{\hbar^{2}}{4
s_{x}t_{x}s_{y}t_{y}}\frac{d\epsilon_{x}d\epsilon_{y}}{\surd(1-
(\epsilon_{x}/(2t_{x}))^{2}\surd 1-(\epsilon_{y}/(2t_{y}))^{2}}
\end{equation}
Using the expressions for the particle energy along the lattice axes,the
exponential factor in eq.(1) can be expressed as
\begin{equation}
\label{(5)} \exp(\beta(\mu-\epsilon))=\exp(\beta(\mu-(\epsilon_{x}+\epsilon_{y})))
\end{equation}
Inserting eqs.(4) and (5) into eq.(1),after some algebra leads to
\begin{eqnarray}
n = \nonumber
\frac{1}{8\pi^{2}s_{x}t_{x}s_{y}t_{y}}\int\int\frac{d\epsilon_{x}d\epsilon_{y}}{1+exp(\beta(\epsilon_{x}+\epsilon_{y}-\mu))}
\nonumber \\
\frac{1}{\surd(1-(\epsilon_{x}/(2t_{x})^{2})}
\frac{1}{\surd(1-(\epsilon_{y}/(2t_{y})^{2}))}
\end{eqnarray}

which is the required equation linking various measurable
parameters of the electron gas and the lattice  with the chemical
potential of the electron gas.The limits of integration are
$\pm2t$ along each of the lattice axes,and the problem is how to
solve this integral,which can not be done in closed analytical
form.Looking at things in a purely mathematical way,the easiest
way to solve this integral is to develop the sub-integral function
into series. Reasoning in this way,the Fermi distribution function
can be re-expressed as follows,under the condition that $\epsilon_{x}+\epsilon_{y}\prec\mu$: 

\begin{equation}
\label{(7)}
\frac{1}{1+exp(\beta(\epsilon_{x}+\epsilon_{y}-\mu))}=\sum_{l=0}^{l=\infty}(-1)^lexp(\beta
l(\epsilon_{x}+\epsilon_{y}-\mu))
\end{equation}

Introducing this development into eq.(6),and slightly grouping the
terms in it,one gets the following expression:
\newpage
\begin{eqnarray}
n=\nonumber
\frac{1}{8\pi^{2}s_{x}t_{x}s_{y}t_{y}}\sum_{l=0}^{l=\infty}(-1)^{l}\int\int
exp(\beta l (\epsilon_{x}+\epsilon_{y}-\mu))\nonumber\\
\frac{d\epsilon_{x}d\epsilon_{y}}{\surd
(1-(\epsilon_{x}/2t_{x})^{2})\surd(1-(\epsilon_{y}/2t_{y})^{2})}
\end{eqnarray}

Integrating this expression,within the limits of $\pm2t$ along
both axes, and performing some algebra,leads to the following
final result for the filling factor of a Fermi gas on a two
dimensional lattice:
\begin{equation}
n=\nonumber \frac{1}{2s_{x}s_{y}}
\sum_{l=0}^{\infty}(-1)^{l}exp(-\beta\mu
l)I_{0}(2lt_{x}\beta)I_{0}(2lt_{y}\beta)
\end{equation}
where the symbol $I_{n}(x)$ denotes the modified Bessel function of
the first kind of the order $n$ and $\beta$ is the inverse temperature.The functions $I_{n}(x)$ are defined by:
\begin{equation}
I_{n}(x)=\sum_{k=0}^{k=\infty}\frac{(x/2)^{n+2k}}{k!\Gamma(n+k+1)}
\end{equation}

\begin{center}
Discussion
\end{center}

Equation (9) represents the general result of this paper,expressed in its most compact form.The obvious question is the applicability of this result to real physical systems. Limiting the summation in eq$(9)$ to terms with $l\le2$ gives 
\begin{eqnarray}
n=\frac{1}{2s_{x}s_{y}}[1-\exp(-\mu/T)I_{0}(\frac{2t_{x}}{T}))I_{0}\frac{2t_{y}}{T}+\nonumber\\ \exp(-2 \mu/T)I_{0}(\frac{4t_{x}}{T})I_{0}(\frac{4t_{y}}{T})]
\end{eqnarray}

Solving this equation for $\exp(-\mu/T)$ leads to 
\newpage
\begin{eqnarray}
exp(-\mu/T)= (I_{0}(\frac{2t_{x}}{T})I_{0}(\frac{2t_{y}}{T})\pm\nonumber\\\sqrt{I_{0}^{2}(\frac{2t_{x}}{T})I_{0}^{2}(\frac{2t_{y}}{T})-4(1-2ns_{x}s_{y})I_{0}(\frac{4t_{x}}{T})I_{0}(\frac{4t_{y}}{T})})/(2I_{0}(\frac{4t_{x}}{T})I_{0}(\frac{4t_{y}}{T})) 
\end{eqnarray}

Developing this equation into series up to second order terms in $t_{x}$ and $t_{y}$ as small parameters,gives the following expression for $exp(-\mu/T)$: 

\begin{eqnarray}
exp(-\mu/T)\cong\frac{1}{2}(1-\sqrt{8ns_{x}s_{y}-3})+\frac{1}{2}[
(t_{x}/T)^{2}+(t_{y}/T)^{2}]\nonumber\\(\frac{16ns_{x}s_{y}-5}{\sqrt{8ns_{x}s_{y}-3}}-3)+\frac{1}{2}(\frac{t_{x}t_{y}}{T^{2}})^{2}(9+\frac{16ns_{x}s_{y}(9-16ns_{x}s_{y})-17}{(8ns_{x}s_{y}-3)^{3/2}})+.\end{eqnarray}

Compare this result with [7] where an expression for the chemical potential of a 1D electron gas has been derived.It follows from eq.(13) that there exists a lower   bound on the values of $n$ for which it is applicable.The condition for validity of eq.(13) is $8ns_{x}s_{y}-3\succ0$ which implies $n\succ\frac{3}{8s_{x}s_{y}}$.The implication of this limitation is that  eq.(13) can not be used for extremely small values of $n$. Inspection of this expression shows that it represents the dependence of the chemical potential of a 2D electron gas on various measurable parameters of the system.As the lattice constants along both lattice axes are taken into account,it follows that this equation gives the possibility of analyzing  the influence of high external pressure on the chemical potential. This possibility could turn out as being useful in calculations of the pressure dependence of the electrical conductivity of 2D organic conductors.

An interesting problem is the value of $n$ (the filling factor) for which the chemical potential becomes equal to zero.It follows after some algebra that $\mu=0$ for $n=n_{0}$,where  

\begin{eqnarray}
n_{0}=\frac{7T^{4}-T^{2}[1+16(t_{x}^{2}+t_{y}^{2})]+32(t_{x}^{2}+t_{y}^{2})^{2}}{16s_{x}s_{y}(T^{2}-2(t_{x}^{2}+t_{y}^{2}))^{2}}\nonumber\\\sqrt{T^{4}+10T^{2}(t_{x}^{2}+t_{y}^{2})+(t_{x}^{2}+t_{y}^{2})^{2}}
\end{eqnarray}

\newpage
The outright implication of this result is that the Lieb-Wu theorem [8] is not generally applicable to 2D systems. 
It can be applied only for certain values of various parameters of the system which enter into eq.(14).It also follows from the last equation that $n_{0}$ is a function of the external pressure to which the specimen is subdued. 
\begin{center}
Conclusions
\end{center} 

In this paper we have determined the chemical potential of the  electron gas on a 2D rectangular lattice.A general expression linking the filling factor,the chemical potential and various measurable parameters of this system was obtained.Starting from this general expression,an approximate expression for the chemical potential has been obtained. It turns out that this expression is much more complicated than the corresponding expression for a 1D electron gas.This expression can be applied in analyzing real experimental data,and in calculations of the electrical  conductivity of Q2D organic metals,which will be done in future work.   
\begin{center}
Acknowledgement
\end{center}

This work was performed as a part of the project 1231 financed by the MNTRS in
Beograd.The purchase of one of the references (J.Low Temp.Phys) was made
possible by a donation of the Royal Dutch Embassy in Yugoslavia.

\begin{center}
References
\end{center}

\leftline{$\left[1\right]$ M.H\'eritier,preprint cond-mat/0111101 (2001).}

\leftline{$\left[2\right]$ P.W.Anderson,preprint cond-mat/0201429 (2002).} 

\leftline{$\left[3\right] $ J.Torr\'es,Ann.Physique(Fr.),{\bf 27},3 (2002).} 

\leftline{$\left[4\right]$ T.Ishiguro,K.Yamaji and G.Saito,{\it Organic
Superconductors}}

Springer Verlag,Heidelberg,(1998).

\leftline{$\left[5\right]$ R.W.Hill,C.Proust,L.Taillefer et.al.,Nature,{\bf
414},711 (2001).}

T.G\"otzfried,A.Weber,K.Heuser et. al.,J.Low Temp.Phys.,{\bf 127},51 (2002).

\leftline{$\left[6\right]$ R.P.Feynman,{\it Statistical Mechanics:A Set of
Lectures}}

W.A.Benjamin Inc.,London (1976).

\leftline{$\left[7\right]$ V.Celebonovic,Phys.Low-Dim.Struct.,{\bf 11/12},25
(1996).}

\leftline{$\left[8\right]$ E.H.Lieb and F.Y.Wu,Phys.Rev.Lett.,{\bf 20},1445
(1968).} and preprint cond-mat/0207529 (2002).

\end{document}